\documentclass[sigconf]{acmart}

\usepackage{booktabs} 

\usepackage[ruled,vlined,linesnumbered]{algorithm2e} 
\usepackage{subcaption}

\acmArticle{4}

\begin{document}

\copyrightyear{2018}
\acmYear{2018} 
\setcopyright{iw3c2w3}
\acmConference[WWW 2018]{The 2018 Web Conference}{April 23--27, 2018}{Lyon, France}
\acmBooktitle{WWW 2018: The 2018 Web Conference, April 23--27, 2018, Lyon, France}
\acmPrice{}
\acmDOI{10.1145/3178876.3186027}
\acmISBN{978-1-4503-5639-8/18/04}

\title{A Coherent Unsupervised Model for Toponym Resolution}

\author{Ehsan Kamalloo}
\affiliation{%
  \institution{University of Alberta}
  \city{Edmonton} 
  \state{Alberta, Canada} 
}
\email{kamalloo@cs.ualberta.ca}

\author{Davood Rafiei}
\affiliation{%
  \institution{University of Alberta}
  \city{Edmonton} 
  \state{Alberta, Canada} 
}
\email{drafiei@cs.ualberta.ca}






\renewcommand{\shortauthors}{E. Kamalloo and D. Rafiei}

\begin{abstract}
Toponym Resolution, the task of assigning a location mention in a document to a geographic referent (i.e., latitude/longitude), plays a pivotal role in analyzing location-aware content. However, the ambiguities of natural language and a huge number of possible interpretations for toponyms constitute insurmountable hurdles for this task. In this paper, we study the problem of toponym resolution with no additional information other than a gazetteer and no training data. We demonstrate that a dearth of large enough annotated data makes supervised methods less capable of generalizing. Our proposed method estimates the geographic scope of documents and leverages the connections between nearby place names as evidence to resolve toponyms. We explore the interactions between multiple interpretations of mentions and the relationships between different toponyms in a document to build a model that finds the most coherent resolution. Our model is evaluated on three news corpora, two from the literature and one collected and annotated by us; then, we compare our methods to the state-of-the-art unsupervised and supervised techniques. We also examine three commercial products including Reuters OpenCalais, Yahoo! YQL Placemaker, and Google Cloud Natural Language API. The evaluation shows that our method outperforms the unsupervised technique as well as Reuters OpenCalais and Google Cloud Natural Language API on all three corpora; also, our method shows a performance close to that of the state-of-the art supervised method and outperforms it when the test data has 40\% or more toponyms that are not seen in the training data.
\end{abstract}

%
%



\maketitle

\section{Introduction}

The size of the Web has been growing near-exponentially over the past decade with a vast number of websites emerging on a variety of subjects and large volumes of textual data being made available every day. In particular, a staggering amount of Web content (such as news articles, blog, forum posts, and tweets) that are added online on a minute by minute basis make frequent use of location names as points of reference. However, many place names have multiple interpretations and using them as references introduces ambiguity which in turn leads to uncertainty.
Determining geographic interpretations for mentions of place names, known as \textit{toponyms}, involves resolving multiple types of ambiguities. 
Toponym resolution is the task of disambiguating or resolving toponyms in natural language contexts to geographic locations (i.e., the corresponding lat/long values). One of the formidable challenges is therefore related to resolving the ambiguity of place names. For example, consider the word \textit{Paris} in the following sentences:

\begin{enumerate}
    \item ``The November 2015 \textit{Paris} attacks were the deadliest in the country since World War II.'' \footnote{\url{https://en.wikipedia.org/wiki/November_2015_Paris_attacks}}
    \item ``\textit{Paris} was voted `the Prettiest Little Town in Canada' by Harrowsmith Magazine.'' \footnote{\url{http://www.brant.ca/en/discover-brant/paris.asp}}
\end{enumerate}

The first sentence cites the tragic incidents in \textit{Paris, France} while in the second sentence, the co-occurrence of \textit{Canada} and \textit{Paris} helps us identify \textit{Paris}. This example illustrates that a toponym resolution method should probe for such clues in documents to reduce the inherent ambiguities of the natural language text. 
GeoNames\footnote{\url{http://geonames.org/}}, the largest crowd-sourced location database, lists 97 interpretations for the place name \textit{Paris}.


The problem of toponym disambiguation has been studied in the literature. Early works on geotagging documents rely on hand-crafted rules and heuristics (e.g., Web-a-Where \cite{Amitay:2004:WGW:1008992.1009040}). Recent studies, however, are grounded on supervised and unsupervised models that do not warrant any manual rules \cite{speriosu-baldridge:2013:ACL2013,Melo:2015:GTD:2837689.2837690,Adelfio:2013:GTR:2525314.2525321,DeLozier:2015:GTR:2886521.2886652,Lieberman:2012:ACF:2348283.2348381}. Adaptive Context Features (or \textit{Adaptive} in short), proposed by \citeauthor{Lieberman:2012:ACF:2348283.2348381} \cite{Lieberman:2012:ACF:2348283.2348381}, and TopoCluster, suggested by \citeauthor{DeLozier:2015:GTR:2886521.2886652} \cite{DeLozier:2015:GTR:2886521.2886652}, are among the prominent methods that have been proposed in this area.
Adaptive method casts toponym resolution as a classification problem, whereas TopoCluster leverages geographical measures to estimate geographical profiles for words.

In this paper, we propose an unsupervised model to tackle toponym resolution since supervised methods yield a poor performance due to the paucity of sufficient annotated data.
Our methods rely merely on the document content and a gazetteer primarily because supplementary information about a Web document often is neither available nor reliable. Clearly, any additional data such as the hosting site of the document and its location (if available) can further improve the performance. 

Our toponym resolution model utilizes context-related features of documents. First, we develop a probabilistic model, called \textit{Context-Bound Hypotheses (CBH)}, inspired by the work of \citeauthor{Rafiei:2016:GNE:2983323.2983795} \cite{Rafiei:2016:GNE:2983323.2983795}, to incorporate two context-related hypotheses into toponym resolution. \citeauthor{Rafiei:2016:GNE:2983323.2983795}'s model aims at geotagging non-location entities and employs a primitive disambiguation technique to spatially resolve toponyms. We extend this model by integrating geographical information of locations into the hypotheses. These context-related premises capture some of the implicit relationships that hold between place names mentioned in the same document; thus, each toponym follows either the location of a frequent toponym or a nearby toponym.
Then, we develop another model, called \textit{Spatial-Hierarchy Sets (SHS)}, which discovers a minimal set of relationships (as discussed in Section~\ref{sec:ProblemDefinition}) that can exist among toponyms. SHS maps the minimality problem to a conflict-free set cover problem wherein sets are constructed using containment and sibling relationships among toponyms. The final model, \textit{Context-Hierarchy Fusion (CHF)}, merges CBH and SHS to exploit context features in extracting minimal relationships.

We conduct extensive experiments to evaluate our model. Our experiments are carried out on multiple datasets, one collected and annotated by us and two others well-known and used in the literature, covering a large range of news sources. We assess the performance of our model and compare it with the state-of-the-art supervised and unsupervised techniques as well as a few commercial geotagging products including Yahoo! YQL Placemaker\footnote{\url{https://developer.yahoo.com/yql/}}, Thomson Reuter's OpenCalais\footnote{\url{http://www.opencalais.com/}}, and Google Cloud Natural Language API\footnote{\url{https://cloud.google.com/natural-language/}}. Moreover, we study the generalization problem of supervised methods by feeding unseen data to Adaptive classifier, showing that the classifier cannot keep up with our unsupervised model.

In summary, the key contributions of this work are as follows:
\begin{itemize}
    \item We devise an unsupervised toponym resolution model that leverages context features of documents as well as spatial relationships of toponyms to produce a coherent resolution.
    \item We extensively evaluate our model on different datasets and in comparison with state-of-the-art methods.
    \item We demonstrate that our unsupervised model surpasses the state-of-the-art unsupervised technique, TopoCluster \cite{DeLozier:2015:GTR:2886521.2886652}, and that it can handle unknown data better than supervised techniques.
\end{itemize}

The rest of this paper is structured as follows: Section~\ref{sec:ProblemDefinition} demonstrates a formal definition of the problem. The proposed unsupervised model is described in Section~\ref{sec:ProposedMethod}. Section~\ref{sec:Experiments} explains the evaluation criteria, gold corpora and experiments. Section~\ref{sec:RelatedWorks} reviews the related work from the literature, and finally, in Section~\ref{sec:Conclusion}, we conclude with a summary of results.

\section{Problem Definition}
\label{sec:ProblemDefinition}
Given a document $\mathcal{D}$ 
and a sequence of toponyms $T=t_1, t_2, \cdots, t_K$ mentioned in $\mathcal{D}$ (e.g., extracted using a named-entity recognizer), toponym resolution
refers to grounding each toponym $t_i$ to a geographic   footprint $\ell_i$ with a latitude and a longitude.

Geographic footprints or references are often derived from a gazetteer, a repository of georeferenced locations and their associated metadata such as type/class, population, spatial hierarchy, etc. 
Following previous works \cite{Lieberman:2012:ACF:2348283.2348381, speriosu-baldridge:2013:ACL2013}, we select GeoNames as our gazetteer primarily because not only is it the largest public location database with sufficiently high accuracy \cite{ahlers2013assessment}, but it also stores the spatial hierarchy of locations\footnote{OpenStreetMap, another well-known crowd-sourced gazetteer, is ruled out since it does not contain spatial hierarchies \cite{gelernter2013automatic}.}. 
Additionally, the bounding boxes of some locations can be retrieved from GeoNames.

Each toponym $t_i$ in $\mathcal{D}$ has a set of location interpretations $L_i = \{l_{i,1}, l_{i,2}, \cdots, l_{i,n_i}\}$, derived from a gazetteer $\mathcal{G}$, where $n_i$ indicates the number of interpretations for toponym $t_i$. Hence, toponym resolution can be seen as detecting a mapping from location mentions $T$ to location interpretations. The resolution method yet cannot enumerate all possible combinations of interpretations. For instance, in a document that contains only 6 U.S. states: \textit{Washington} ($n_1$=113), \textit{Florida} ($n_2$=228), \textit{California} ($n_3$=225), \textit{Colorado} ($n_4$=230), \textit{Arizona} ($n_5$=63) and \textit{Texas} ($n_6$=53), the number of possible interpretations exceeds 4 billion. 
The past works in this area therefore incorporate heuristics to reduce the immense search space. For instance, picking the most populated interpretation is a simple heuristic that has been adopted in early works \cite{Leidner:2007:TRT:1328964.1328989}. However, population alone cannot be effective for an off-the-shelf resolution system. We address this problem by looking into containment and sibling relationships among toponyms in a document.

\section{The Unsupervised Model}
\label{sec:ProposedMethod}
The proposed method leverages a combination of context-related features of documents to address toponym resolution. These features are grounded on the characteristics of toponyms. It is well-accepted (e.g., SPIDER \cite{speriosu-baldridge:2013:ACL2013}) that toponyms mentioned in a document often show the following minimality properties:

\begin{itemize}
    \item \textit{one-sense-per-referent}: all of the occurrences of a toponym generally refer to a unique location within a single document;
    \item \textit{spatial-minimality}: toponyms mentioned in a text tend to be in a spatial proximity of each other.
\end{itemize}

In this section, we develop context-bound hypotheses, inspired by the named entity geotagging method suggested by \citeauthor{Rafiei:2016:GNE:2983323.2983795} \cite{Rafiei:2016:GNE:2983323.2983795}. Then, we describe spatial hierarchies built from containment and sibling relationships among location mentions in text. Lastly, we explain how these two methods coalesce into an unsupervised model to disambiguate toponyms.

\subsection{Context-Bound Hypotheses}
\label{sec:CBH}
\citeauthor{Rafiei:2016:GNE:2983323.2983795} \cite{Rafiei:2016:GNE:2983323.2983795} propose a probabilistic model to associate named entities to locations. The task of geotagging named entities is delineated as follows: given a named entity and a set of documents germane to it, a geotagger finds the geographic focus of the named entity. The model, introduced by \citeauthor{Rafiei:2016:GNE:2983323.2983795} \cite{Rafiei:2016:GNE:2983323.2983795}, incorporates two hypotheses: {\em geo-centre inheritance hypothesis} and {\em near-location hypothesis} and estimates the probabilities that these premises hold. The probabilistic model makes use of the known entities that are mentioned in the surrounding text to determine the geo-centre of a named entity. Their geotagging task mainly focuses on non-location named entities and does only a simple location disambiguation on each toponym, independent of other toponyms in the same document. A question here is if their probabilistic model can be applied to toponym resolution. This is the question we study in our {\em Context-Bound Hypotheses (CBH)} model. In particular, to model the cohesion of toponyms to context, 
we integrate the hypotheses with geographical information of locations in order to spatially locate a place mention. Context-Bound assumptions allow us to reduce toponym resolution to a probabilistic model, which we are set to compute the estimations in this section.

\begin{algorithm}
\DontPrintSemicolon

\caption{Preliminary Toponym Disambiguation in CBH}
\label{alg:PreliminaryDisambiguation}

\KwIn{Document $\mathcal{D}$ and sequence of toponyms $T$}
\KwOut{A preliminary mapping from $T$ to location interpretations}

\SetKw{In}{in}

\SetKwFunction{FRetrieveHierarchy}{RetrieveHierarchy}
\SetKwFunction{FLookUp}{LookUp}
\SetKwFunction{FMentions}{Mentions}
\SetKwFunction{FTD}{TD}

$resolution \gets \emptyset$ \;
\For{{\upshape toponym} $t_i$ \In $T$}{
    $\ell_i \gets \textsc{Nil}$ \;
    \BlankLine
    
    \For{{\upshape interpretation} $l_{i,j}$ \In $L_i$}{
    $\mathcal{H}_{i,j} \gets $ \FRetrieveHierarchy{$l_{i,j}$} \;
    \BlankLine
    $node \gets $ \FLookUp{$parent[l_{i,j}], \mathcal{H}_{i,j}$} \;
    
    $score \gets 0$ \;
    
    \While{$node \neq$ {\sc Nil}}{
        \For{$m_h$ \In \FMentions{$node$}} {
            
            \For{$m_l$ \In \FMentions{$l_{i,j}$}} {
                $similarity \gets \max(similarity, \frac{1}{\FTD{$m_h, m_l$}})$
            }
            
            $score \gets score + similarity$ \;
        }
        
        $node \gets parent[node]$ \;
    }
    
    \uIf{$confidence[\ell_i] < score$} {
        $\ell_i \gets (l_{i,j}, score)$ \;
    }
    \ElseIf{$confidence[\ell_i] = score$} {
        \If{$population[\ell_i] < population[l_{i,j}]$} {
            $\ell_i \gets (l_{i,j}, score)$ \;
        }
    }
  }
  
  $resolution \gets resolution \cup (t_i, \ell_i)$ \;
}

\end{algorithm}

The space of possible interpretations (as shown with an example of 6 U.S. states) can be huge and enumerating all combinations may not be feasible.
To be able to compute probabilities of the hypotheses, we perform a preliminary location disambiguation \cite{Rafiei:2016:GNE:2983323.2983795}. This procedure, which is shown in Algorithm \ref{alg:PreliminaryDisambiguation}, leverages a heuristic to resolve toponyms. Consider a location interpretation $l_{i,j}$ of toponym $t_i$. The mentions of the ancestors in $l_{i,j}$'s spatial hierarchy (line 5; the hierarchies can be obtained from gazetteer $\mathcal{G}$) can be used as clues to resolve toponym $t_i$. The closer an ancestor mention is, the more chance that particular interpretation has to get selected. For example, toponym {\em Edmonton} refers to 6 different locations. Provided that it co-occurred with either {\em Alberta} or {\em Canada}, we can pinpoint it (i.e., the city of {\em Edmonton} located in {\em Canada}). For each toponym $t_i$, the preliminary disambiguation measures a score for each interpretation $l_{i,j}$ (lines 8-13) and picks the interpretation with maximum score (lines 14-15) and in case of tie, the most populous interpretation is selected (lines 16-18). The score is acquired by finding the maximum similarity between $l_{i,j}$ mentions and its ancestors' mentions; similarity here is the inverse of term distance (line 11), as used by \citeauthor{Rafiei:2016:GNE:2983323.2983795} \cite{Rafiei:2016:GNE:2983323.2983795}.

Preliminary disambiguation works poorly in cases where no mentions of locations in spatial hierarchy exist in the document. For instance, suppose we find toponyms {\em Toronto}, {\em London}, and {\em Kingston} in an article. Though, humans can recognize that these cities are presumably located in {\em Ontario, Canada}, preliminary resolution is unable to find any clues for disambiguation and as a result, assigns the toponyms to the interpretation with the highest population (i.e., {\em Toronto} $\mapsto$ {\em Canada}, {\em London} $\mapsto$ {\em England}, and {\em Kingston} $\mapsto$ {\em Jamaica}).

The result of the initial phase can be augmented by incorporating context-related features into the resolution process. Our CBH model proceeds to compute probabilities for the two hypotheses. The method operates at each administrative division separately since toponyms may lie in disparate division levels. Hence, the method begins the disambiguation process from the lowest division and furthers the process until all toponyms are resolved.

The geo-centre inheritance indicates that the location interpretation of a toponym can be drawn from the geographic scope of the document. The entities (i.e., people, locations, and organizations) used in an article, can ascertain a location interpretation to which the article is geographically relevant \cite{Andogah:2012:DGS:2388118.2388201}. This location defines the geographic scope of the document.

Based on the inheritance hypothesis, the toponyms mentioned in a document are more likely to be part of or under the same administrative division as the geographic scope of the document. This makes sense due to the {\em spatial minimality} property. Therefore, we first estimate the geographic scope of the document via a probabilistic model. In particular, for toponym $t_i$ at division $d$, the probability of $l_{i,j}$ being the correct interpretation is
\begin{equation}
P_{\mathrm{inh}}^{(d)}( l_{i,j} | \mathcal{D}, t_i ) = \frac{\mathrm{tf} \: \left( \: \mathrm{anc}_d \: (l_{i,j}) \:\right)}{\sum_{p = 1}^{n_i} \mathrm{tf} \: \left( \: \mathrm{anc}_d \: (l_{i,p}) \: \right)}
\label{eq:P_inh}
\end{equation}
where $\mathrm{anc}_d$ returns the ancestor of an interpretation at division $d$ and $\mathrm{tf}(w)$ computes the term frequency in the document.
Each location interpretation here is extended to include its corresponding spatial hierarchy. For example, interpretations of toponym {\em Paris} are represented as
\[
\begin{split}
\bigl\{ \; &[Paris \leadsto Ile\text{-}de\text{-}France \leadsto France], \\
&[Paris \leadsto Lamar County \leadsto Texas \leadsto US], \cdots \; \bigr\}
\end{split}
\]

The second hypothesis, namely near-location hypothesis, relies upon the toponyms mentioned in the vicinity of a toponym. Toponyms nearby a toponym can be linked to one another primarily because of {\em object/container} and {\em comma group} relationships they possibly have \cite{Lieberman:2010:GUP:1722080.1722088}. According to this hypothesis, the closer toponym $s$ to toponym $t$, the stronger evidence toponym $s$ is to disambiguate toponym $t$. This is why, in this hypothesis, we compute the term distance between toponyms as a measure of similarity to estimate probabilities. In effect, for toponym $t_i$ at division $d$, the probability of $l_{i,j}$ being the correct interpretation is
\begin{equation}
P_{\mathrm{near}}^{(d)}( l_{i,j} | \mathcal{D}, t_i ) = \frac{\mathrm{sim} \: \left( t_i \: , \; \mathrm{anc}_d \: (l_{i,j}) \right)}{\sum_{p = 1}^{n_i} \mathrm{sim} \: \left( t_i \: , \; \mathrm{anc}_d \: (l_{i,p}) \right)}
\label{eq:P_near}
\end{equation}
where $\mathrm{sim}( v_1, v_2 )$ is the similarity function between terms $v_1$ and $v_2$ as demonstrated below:
\begin{equation}
\mathrm{sim}(v_1, v_2) = \frac{1}{\min_{w_i \in M(v_i)}\{\text{{\sc TD}}(w_1, w_2)\}}
\end{equation}
where $\mathrm{\textsc{TD}}(w_1, w_2)$ is the distance between indices of $w_1$ and $w_2$ and $M(v)$ is a set containing the mentions of term $v$ in document $\mathcal{D}$.

Now, we combine $P_{\mathrm{inh}}^{(d)}$ and $P_{\mathrm{near}}^{(d)}$ to incorporate both premises into the model. The final context-bound model is regarded as a weighted linear function of the two probabilities:
\begin{equation}
\begin{split}
P_{\mathrm{CB}}^{(d)}( l_{i,j} | \mathcal{D}, t_i ) &= \: J^{(d)}(\mathcal{D}, t_i) \cdot P_{\mathrm{near}}^{(d)}( l_{i,j} | \mathcal{D}, t_i ) \\
 &+ \: (1 - J^{(d)}(\mathcal{D}, t_i)) \cdot P_{\mathrm{inh}}^{(d)}( l_{i,j} | \mathcal{D}, t_i )
\end{split}
\label{eq:P_context}
\end{equation}

The coefficient $J^{(d)}(\mathcal{D}, t_i)$ is obtained via Shannon Entropy of the vector induced by near-location probabilities for toponym $t_i$ with respect to $l_{i,j}$ for all values of $j$. 

The resolution is undertaken through maximum likelihood estimation over the probability in Equation \eqref{eq:P_context}. The final computed probability can be considered as confidence score.

\begin{algorithm}
\DontPrintSemicolon

\caption{CBH Resolution}
\label{alg:CBHProc}

\KwIn{Document $\mathcal{D}$ and sequence of toponyms $T$}
\KwOut{A mapping from $T$ to location interpretations}

\SetKw{In}{in}
\SetKw{To}{to}

\SetKwFunction{FPreliminary}{PreliminaryResol}
\SetKwFunction{Fargmax}{argmax}

$resolution \gets $ \FPreliminary{$\mathcal{D}$, $T$}  \tcp{Alg. \ref{alg:PreliminaryDisambiguation}}

\For{$k \gets 1 \; \To \; maxIterations$}{

    \For{{\upshape division} $d$ \In $\{County, State, Country\}$}{
    
        \For{{\upshape toponym} $t_i$ \In $T$}{
            $\ell_i \gets \Fargmax_j\{P_{\mathrm{CB}}^{(d)}( l_{i,j} | \mathcal{D}, t_i )\}$ \;
            \tcp{refer to Eq. \eqref{eq:P_context}}
            $resolution \gets resolution \cup (t_i, \ell_i)$ \;
        }
    }
}

\end{algorithm}

In summary, the CBH resolution method is illustrated in Algorithm \ref{alg:CBHProc}. The approach starts with a preliminary resolution, followed by a hypotheses assessment to rectify results from the initial resolution. The hypotheses model computes the probabilities for each division separately to ensure the model can afford toponyms in all levels of dispersion. Once the modification process finished, the algorithm repeats for another iteration since altering the resolution of a toponym may affect other disambiguated toponyms. Our experiments show that CBH often takes two iterations to complete. However, in some cases, the modification step never terminates. Specifically, consider the following sentence, an excerpt from a news article:
\begin{quote}
    ``... London's Heathrow, one of the world's busiest travel hubs.'' \footnote{http://money.cnn.com/2016/12/14/news/companies/british-airways-ba-strike-christmas}
\end{quote}
{\em London} and {\em Heathrow} are recognized as toponyms. Because no notion of ancestors in the spatial hierarchy can be found, the initial resolution favors the highest population interpretation (i.e., {\em London} $\mapsto$ {\em England} and {\em Heathrow} $\mapsto$ {\em Florida, US}). In the next step, the hypotheses model maps {\em London} to a place in United States because the other toponym is located in United States. Accordingly, {\em Heathrow} is assigned to the airport in England. After the first iteration, the resolution is changed to $\{London \mapsto US , Heathrow \mapsto England \}$. Conversely, the second iteration would alter the results to $\{London \mapsto England , Heathrow \mapsto US \}$; the algorithm is now trapped in an infinite loop. This is why, we introduce $maxIterations$ parameter to eschew these circumstances. While CBH fails to successfully resolve toponyms in such cases, the approach, described in next section, can address this shortcoming.

\subsection{Spatial-Hierarchy Sets}
\label{sec:SHS}
The spatial minimality property (noted by Leidner \cite{Leidner:2007:TRT:1328964.1328989}) leads us to another resolution method called {\em Spatial-Hierarchy Sets (SHS)}. This method is grounded on containment and sibling relationships that are likely to exist among toponyms in a document. Consider a non-disjoint partitioning of the universe of locations (in a gazetteer) where locations with similar or related interpretations (e.g., those under the same administrative division or within a close  proximity) form a partition. 
Since toponyms in a document tend to refer to geographically related locations, and those locations are more likely to be in the same partitions than different partitions, we want to find a small set of partitions that cover all toponyms; this
can be modeled as a conflict-free covering problem. Conflict-free covering refers to the traditional set cover problem where each element must be covered by at most one set in the answer. The covering needs to be conflict-free due to {\em one-sense-per-referent} property. We formally define conflict-free covering as an instance of the conflict-free coloring of regions \cite{Har-Peled:2003:CCP:777792.777810}.
\subsubsection*{Conflict-free Covering Problem} Given a finite family of finite sets $\mathcal{S}$ where each set $S_i$ is associated with a non-negative weight $w_i$ and a universal set $\mathcal{U}$ containing all the elements from the sets, we seek to find a collection of sets, namely $\mathcal{A}$, with minimum weight such that their union becomes $\mathcal{U}$ while each element is covered by at most one set in $\mathcal{A}$.

We formulate toponym resolution by conflict-free covering problem as the following:

\begin{enumerate}
    \item
    Each parent with all its children form a set of related interpretations. Let $\mathcal{S}$ denote the collection of all such sets that can be constructed. Each parent appears in a set with its children, hence the size of $\mathcal{S}$ is the same as the number of parents with non-zero children. Algorithm~\ref{alg:SetGeneration} depicts the details of generating $\mathcal{S}$.
    
    \item \label{item:Step2}
    Recall that $T$ denotes the set of toponyms in document $\mathcal{D}$ (as defined in  Section~\ref{sec:ProblemDefinition}). We say a set in $\mathcal{S}$ covers a toponym in $T$, if the set contains the surface text of the toponym. We want to select sets in $\mathcal{S}$ that cover all toponyms in $T$. Our goal is to minimize the number of interpretations (spatial minimality) by selecting as few sets in $\mathcal{S}$ as possible.
    
    \item
    Let us form a color class for each toponym. The color class for a toponym includes all possible interpretations of the toponym. For example, {\em Texas} is a color class which includes all places that can resolve {\em Texas}. We want to avoid selecting multiple interpretations for the same toponym. That means, we do the selection in (\ref{item:Step2}) with the constraint that no more than one color or interpretation can be selected for each toponym.
    \end{enumerate}

\begin{algorithm}
\DontPrintSemicolon

\caption{Spatial-Hierarchy Set Generation}
\label{alg:SetGeneration}

\KwIn{Document $\mathcal{D}$ and sequence of toponyms $T$}
\KwOut{$\mathcal{S}$, a collection of spatial hierarchy sets}

\SetKw{In}{in}
\SetKw{False}{false}
\SetKw{True}{true}
\SetKw{Skip}{skip}

\SetKwFunction{FAddChild}{AddChild}

$\mathcal{S} \gets \emptyset$ \;
$P \gets \emptyset$
\BlankLine
\For{{\upshape toponym} $t_i$ \In $T$} {
    \If{$name[t_i]$ \In $P$} {
        \Skip $t_i$
    }
    
    \For{{\upshape interpretation} $l_{i,j}$ \In $L_i$} {
        \tcp{checks whether the set exists}
        \uIf{$parent[l_{i,j}]$ \textbf{in} $\mathcal{S}$} {
            \FAddChild{$(l_{i,j}, \True), \mathcal{S}[parent[l_{i,j}]]$}
        }
        \Else {
            $\mathcal{S} \gets \mathcal{S} \cup \{(parent[l_{i,j}], \False) \rightarrow (l_{i,j}, \True) \}$ \;
            \tcp{the new set is a tree rooted at $parent[l_{i,j}]$}
            \tcp{the boolean values represent $mentioned$ flags}
            \If{$l_{i,j}$ \In $\mathcal{S}$} {
                $mentioned[\mathcal{S}[l_{i,j}]] \gets \True $
            }
        }
        
        $P \gets P \cup name[l_{i,j}]$
  }
}

\end{algorithm}

In the special case where the color classes are empty (i.e., no constraint on colors), the problem becomes the classic set cover, which is NP-complete. This means that existing methods approximate the optimal solution. We leverage a greedy approach \cite{Slavik:1996:TAG:237814.237991} to solve the problem. Although the greedy approach gives an approximate answer to the problem in general, our experiments reveal that such answer yield a competitive performance.

However, this model suffers from some deficiencies, even if an optimal solution is reached. A problem with this formulation is that we cannot have {\em Montreal, Quebec} and {\em Windsor, Ontario} in the same text (or they will not be resolved correctly) because {\em Windsor} is also a town in {\em Quebec}. These are cases where the hypotheses model, namely CBH, can better resolve. Furthermore, there may be circumstances that similar toponyms may appear in more than one sets and yet, we cannot favor one set to another. Suppose we have a document where only {\em Georgia} and {\em Turkey} are mentioned.
Two sets, $\{Georgia\mathrm{(city)}\leadsto Texas \mathrm{(state)}, \; Turkey \mathrm{(city)} \leadsto Texas \mathrm{(state)}\}$ and $\{Georgia \mathrm{(country)}\leadsto World, \; Turkey \mathrm{(country)} \leadsto World\}$, would emerge in $\mathcal{S}$. Without any additional information, such as document source, even humans cannot choose the correct interpretation. SHS selects the most populated set as a rule of thumb in these cases.


\subsection{Context-Hierarchy Fusion}
\label{sec:CHF}
While the Spatial-Hierarchy Sets approach guarantees the minimality properties, it fails to select between identical structures
(e.g., the {\em Georgia} and {\em Turkey} case)
mostly because it does not delve into other context-related features of the document. On the other hand, the Context-Bound Hypotheses model benefits from term frequency and term distance features of the context. Notwithstanding the situations like {\em Georgia} and {\em Turkey}, using other context sensitive information alleviates the disambiguation process in most cases. For example, toponyms {\em London}, {\em Aberdeen} and {\em Edinburgh} have interpretations located in {\em Canada} and SHS resolves them to the corresponding interpretations in {\em Canada} to preserve minimality. Even the presence of toponym {\em England} does not change the result because {\em Aberdeen} and {\em Edinburgh} located in {\em Scotland} and we still need to pick two sets to attain the correct resolution.

Consequently, merging SHS and CBH method allows us to take advantage of both methods at the same time. {\em Context-Hierarchy Fusion (CHF)} method chooses an interpretation from CBH only if the confidence score is higher than a threshold $\tau$. Otherwise, it resolves toponyms using SHS.

\begin{table}[t]
\begin{center}
\begin{tabular}{l c c c} 
 \hline
  & TR-News & LGL & CLUST \\ 
 \hline
 News sources & 36 & 85 & 352 \\
 Documents & 118 & 588 & 13327 \\ 
 Annotated docs & 118 & 588 & 1082 \\
 Annotated topos & 1318 & 5088 & 11962 \\
 Topos with GeonameID & 1274 & 4462 & 11567 \\
 Distinct topos & 353 & 1087 & 2323 \\
 Median topos per doc & 9 & 6  & 8 \\
 Topos not found in GeoNames & 2.7\% & 3.2\% & 3.3\% \\
 Wikipedia-linked topos & 94.3\% & 94.1\% & 94.2\% \\
 \hline
\end{tabular}
\caption{Corpora used in our experiments}
\label{tab:datasets}
\end{center}
\end{table}

\section{Experiments}
\label{sec:Experiments}
In this section, we conduct extensive experiments to evaluate our methods\footnote{The source code and the annotated dataset is available at \url{https://github.com/ehsk/CHF-TopoResolver}} and to assess their performance under different settings. The particular questions to be investigated are:
\begin{enumerate}
    \item Given that CBH comprises different steps and components, how much does inheritance and near location hypothesis improve upon the preliminary location disambiguation?
    \item How sensitive is Context-Hierarchy Fusion to the value of the threshold and if there are some sweet spots?
    \item How accurate is the proposed method, compared to the state-of-the-art supervised and unsupervised methods as well as commercial systems?
    \item How does the proposed method compare to the state-of-the-art supervised method in terms of the generality of the model on unseen data?
    \item When is an unsupervised technique expected to surpass supervised methods?
\end{enumerate}
For (3), we compare the performance of our method to that of the state-of-the-art methods as well as commercial systems (i.e., Yahoo! YQL Placemaker, OpenCalais and Google Cloud Natural Language API). The details of these proprietary products have not been made public. However, these systems can be accessed  through public Web APIs at a relatively liberal rate limit, which enable us to automatically test their geotagging process on our datasets.


In our evaluation setting, we apply two methods for toponym recognition. First, we assume that the recognition phase is flawless, which is displayed as \textit{Resol}. In this method, the annotated toponyms without latitude/longitude are fed to the underlying resolution method. These experiments are conducted to compare our methods to resolution methods such as TopoCluster \cite{DeLozier:2015:GTR:2886521.2886652}. Second, we employ Stanford NER
\cite{Finkel:2005:INI:1219840.1219885} to tag locations, which is shown by \textit{GeoTag}.
We run \textit{GeoTag} experiments to draw a comparison with systems performing both recognition and resolution including closed-source products and Adaptive \cite{Lieberman:2012:ACF:2348283.2348381}.

\subsection{Datasets}
In order to evaluate our toponym resolution methods, gold data corpora are required, in which all occurrences of geographic names and phrases have been manually annotated. In our experiments, we exploit three annotated datasets. Table~\ref{tab:datasets} summarizes and compares the statistics of these datasets.

\subsubsection*{TR-News}
We collected this dataset from various global and local News sources. We obtained news articles from several local news sources to include less dominant interpretations of ambiguous locations such as {\em Edmonton, England} and {\em Edmonton, Australia} rather than {\em Edmonton, Canada} or {\em Paris, Texas, US} in lieu of {\em Paris, France}. Additionally, a number of articles from global news sources such as BBC and Reuters have been selected to preserve the generality of the corpus. We manually annotated toponyms in the articles with the corresponding entries from GeoNames. The gold dataset consists of 118 articles.

\subsubsection*{Local-Global Lexicon (LGL)}
This corpus was developed by \citeauthor{5447903} \cite{5447903}. It is collected from local news sources and mainly focuses on including ambiguous toponyms and this is why, it is suitable to test toponym resolution systems against geographically localized documents. The dataset is composed of 588 articles from 85 news sources.

\subsubsection*{CLUST}
\citeauthor{Lieberman:2011:MTR:2009916.2010029} \cite{Lieberman:2011:MTR:2009916.2010029} compiled this dataset from a variety of global and local news sources. CLUST is a large dataset containing 1082 annotated articles. \\

According to Table~\ref{tab:datasets}, the median number of toponyms per document in all datasets are close to each other, meaning that the corpora do not differ significantly with one another in terms of the number of toponyms per article.

In addition, the three datasets contain toponyms (roughly 3\%) that cannot be found in gazetteer $\mathcal{G}$, while annotated and linked to an entry in the gazetteer. We observe that such toponyms fall into one of the following categories: uncommon abbreviations such as {\em Alta.} stands for {\em Alberta, Canada}, multi-word places such as {\em Montreal-Pierre Elliott Trudeau International Airport}, and transliterated place names (e.g., city of {\em Abbasiyeh, Egypt} written as {\em Abbassiya}).

\begin{figure}[t]
\centering
\includegraphics[width=2.83in]{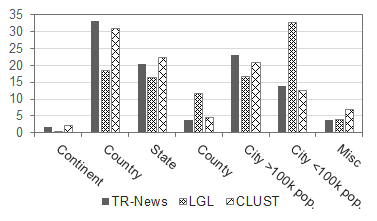}
\caption{Comparative analysis of the test datasets based on location type}
\label{fig:DatasetComparison}
\end{figure}

The test corpora is also analyzed by the location type of their annotated toponyms, as done by \citeauthor{Lieberman:2012:ACF:2348283.2348381} \cite{Lieberman:2012:ACF:2348283.2348381}. We compute the percentage of each location type for each dataset. As show in Figure~\ref{fig:DatasetComparison}, $LGL$ dataset largely consists of small cities, which makes it a challenging test dataset since well-known locations are presumably to be resolved with high precision due to their frequent use in articles. In contrast, $TR$-$News$ and $CLUST$ datasets are roughly similar and include countries more than any location type. This denotes that the articles appeared in $TR$-$News$ and $CLUST$ are extracted from sources that are aimed at a global audience. These sources usually provide more details for location mentions such as saying {\em Paris, US} instead of {\em Paris}. On the other hand, in $LGL$, because the articles are meant to be of use for local audience, the news publishers typically do not state additional information in this regard. Thus, geotagging approaches can be tested against these test corpora since they span a variety of news sources both globally and locally.

\subsection{Evaluation Metrics}
 Performance measures in our experiments are $Precision$, $Recall$, $F_1$-$measure$, and mean error distance ($M$). However, to ascertain whether an interpretation is correctly predicted, we also investigate the error distance between the predicted coordinates and the actual coordinates, as used in numerous studies \cite{Cheng:2010:YYT:1871437.1871535,DeLozier:2015:GTR:2886521.2886652, speriosu-baldridge:2013:ACL2013, Leidner:2007:TRT:1328964.1328989,Santos2015,Lieberman:2012:ACF:2348283.2348381,Ardanuy:2017:TDH:3078081.3078099}. This distance enables us to fare various systems against each other since they may select latitude/longitude of locations from different gazetteers or knowledge bases\footnote{Locations are represented with a single centroid and gazetteers may vary in picking the centroids.}. We set the error distance to 10 miles, same as Adaptive method \cite{Lieberman:2012:ACF:2348283.2348381}, whereas most researches tend to adopt a relaxed threshold (i.e., 161 kilometers) \cite{speriosu-baldridge:2013:ACL2013,DeLozier:2015:GTR:2886521.2886652,Santos2015,Ardanuy:2017:TDH:3078081.3078099}.
 
 For TopoCluster \cite{DeLozier:2015:GTR:2886521.2886652} and the commercial products, on the other hand, we employ a different criteria primarily because the error distance may not be accurate for large areas (even with higher error distance thresholds). Hence, in order to consider whether an interpretation is correctly projected to a coordinate, we check if the predicted interpretation resides in the bounding-box area of the ground truth; here the bounding-boxes of locations are extracted from GeoNames. We did not use this bounding-box grounded accuracy for other methods since most of them rely on the same gazetteer adopted in this work. Although using bound-boxes works in favor of TopoCluster and the proprietary products, the mean error distance fails to precisely mirror the accuracy for these methods since for a prediction deemed as correct based on bounding-boxes, the error distance can still be high.
 
 Furthermore, in \textit{Resol} experiments, we only calculate $Precision$ because given a toponym, a resolution method is more likely to map it to an interpretation unless it does not exist in the gazetteer; thus, $Recall$ would be approximately analogous to $Precision$. It is also worth mentioning that the mean error distance is only reported in \textit{Resol} experiments and not in \textit{GeoTag} experiments, because the mean error distance cannot be measured for toponyms that are either not identified or falsely detected.
 
\subsection{Analysis of Context-Bound Hypotheses}
As discussed in Section~\ref{sec:CBH}, Context-Bound Hypotheses commences with a preliminary toponym disambiguation, followed by estimating two probabilities for inheritance and near-location hypotheses. In this section, we evaluate the preliminary phase and see whether the modification phase by Context-Bound hypotheses alleviates the resolution performance. Moreover, we study the role of the hypotheses in CBH by removing one of them at a time and measuring the performance. This experiment is conducted on the $TR$-$News$ dataset in both \textit{Resol} and \textit{GeoTag} modes.

As shown in Table~\ref{tab:CBH}, taking both hypotheses into account complements the preliminary disambiguation, though the improvement does not seem considerable (slightly higher than 1\% in $F_1$-measure) in both \textit{GeoTag} and \textit{Resol} experiments.

\begin{table}[t]
\begin{center}
 \begin{tabular}{l c c c c} 
 \hline
  & $P_\mathrm{Resol}$ & $P_\mathrm{GeoTag}$ & $R_\mathrm{GeoTag}$ & $F_{1-\mathrm{GeoTag}}$ \\
 \hline
 Preliminary & 78.0 & 73.4	& 52.1 & 60.9 \\
 Inheritance & 78.1 & 73.6	& 52.0  & 61.0	 \\
 Near-location & 79.0 & 73.9  & 52.3 & 61.2 \\
 CBH            & 79.2 & 74.9	& 53.0	& 62.1 \\
 \hline
\end{tabular}
\caption{Detailed analysis of Context-Bound Hypotheses (CBH) on $TR$-$News$ dataset}
\label{tab:CBH}
\end{center}
\end{table}

Additionally, the near-location hypothesis contributes to the improvement more than the inheritance hypothesis. This is largely because the inheritance hypothesis estimates probabilities using term frequency. In cases where two locations are mentioned as frequent as each other, term frequency does not seem accurate. For example, consider the toponym {\em Edmonton}, which can be located in either {\em Canada} or {\em Australia} in a document where {\em Australia} and {\em Canada} appear twice each. This results in the same score for both interpretations and a decision would be made by population size. Term distance, however, can help better in this case, denoting that the closer mention is more likely to be the correct interpretation. Nonetheless, we still need both hypotheses since the results are improved by putting near-location and inheritance together.

\subsection{Fusion Threshold Study}
\begin{figure}[t!]
    \centering
    \includegraphics[width=2.82in]{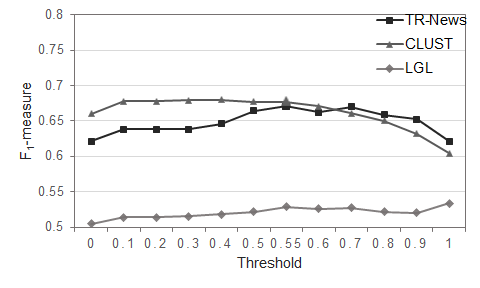}
    \caption{$F1$-$measure$ vs. threshold $\tau$ for Context-Hierarchy Fusion method on $TR$-$News$ dataset. At $\tau=0.55$, CHF achieves the best $F_1$-measure on all three corpora.}
    \label{fig:ThresholdStudy}
\end{figure}

In Context-Hierarchy Fusion (explained in Section~\ref{sec:CHF}), choosing an appropriate value for the threshold can be crucial in the resolution the performance. In this experiment, we vary the threshold $\tau$ to study its effect on performance. According to the results shown in Figure~\ref{fig:ThresholdStudy}, we can identify a sweet spot when CHF achieves the best performance on all three datasets; this happens when $\tau$ falls between 0.5 and 0.6; we set $\tau$ to 0.55 in our experiments.

Also, we can see a mild spike in $F_1$-$measure$ at $\tau=1$ in the $LGL$ curve, which can be attributed to the localized content of the dataset. In particular, SHS (at $\tau=1$, CHF is analogous to SHS) works better on $LGL$ since locations in $LGL$ are not mentioned frequently alongside their corresponding spatial hierarchy ancestors. As discussed in Section~\ref{sec:CBH}, CBH needs to spot the mentions of these ancestors in documents (containment relationship) in order to generate a more accurate resolution, whereas SHS does not rely solely on containment relationships. It also takes sibling relationships into account, and as a result, merging SHS and CBH does not seem to be effective on $LGL$.


\subsection{Resolution Accuracy}
In this section, we measure the performance of our proposed methods and compare them with other resolution techniques. The methods presented in this paper are Context-Bound Hypotheses (CBH), Spatial-Hierarchy Sets (SHS) and Context-Hierarchy Fusion (CHF). We compare the results with two prominent systems: TopoCluster \cite{DeLozier:2015:GTR:2886521.2886652} (i.e., the state-of-the-art unsupervised model) and Adaptive \cite{Lieberman:2012:ACF:2348283.2348381} (i.e., the state-of-the-art supervised model). The source code of TopoCluster was available online, so we were able to test the method on our datasets. However, in order to test the Adaptive classifier, we implemented the supervised method\footnote{Since we did not have access to the source code.}, albeit without two features, namely \textit{dateline} and \textit{locallex}; this was because for \textit{locallex}, the authors used an annotated dataset containing the expected audience location of the news sources and also, \textit{dateline} required a general location for each article which was not available for most articles in the test corpora. The modified version is named {\em CustomAdaptive} in our results. We follow the same parameter setting of the original Adaptive \cite{Lieberman:2012:ACF:2348283.2348381} and perform 10-fold cross validation to test CustomAdaptive.

Table~\ref{tab:resolution} illustrates the evaluation results. CHF produces the best performance among our proposed methods on $CLUST$ and $TR$-$News$ and SHS beats the other proposed techniques on $LGL$. Among all listed methods, CustomAdaptive shows the highest performance. We also report recall, to make a comparison with the original Adaptive method \cite{Lieberman:2012:ACF:2348283.2348381}.

While commercial products produce high precision, their recall is lower than our proposed methods in all cases except for Yahoo! YQL Placemaker. Placemaker yields the best results among the commercial products and achieves higher overall performance than our methods. On the other hand, OpenCalais is able to recognize toponyms as locative expressions. For instance, it identifies {\em the Kenyan captial} rather than just {\em Kenyan}. However, we observe that sometimes it fails to detect a full location phrase; for example, only {\em Toronto} in {\em Greater Toronto Area} is detected\footnote{We count these as correct resolutions unless they fall outside the bounding box of the annotated toponym.}. Further, Google Cloud Natural API offers an entity extraction service, which focuses highly on recognition of named entities\footnote{Google Cloud Natural API extracts locative expressions in any form in addition to proper names like {\em family home} and {\em suburb}.}. The system links extracted entities to their corresponding Wikipedia articles and provides no additional information about geographic coordinates of location entities. Therefore, the geographical information of locations can only be derived from Wikipedia for this product.
According to Table~\ref{tab:datasets}, nearly 94\% of toponyms in each dataset have Wikipedia articles\footnote{GeoNames keeps record of Wikipedia URLs for each location.}, but not all Wikipedia articles contain spatial coordinates of locations, which is partly attributed to a poor recall in our experiments. Thus, we can see why entity linking approaches cannot be exploited for toponym resolution.

We run {\em Resol} experiments to analyze TopoCluster \cite{DeLozier:2015:GTR:2886521.2886652} since it is a resolution method. \citeauthor{DeLozier:2015:GTR:2886521.2886652} stipulated that TopoCluster performs best when integrated with a gazetteer; this is why, the integrated version, called {\em TopoClusterGaz}, is adopted throughout this experiment. The results are presented in Table~\ref{tab:resolution} ($P_\mathrm{Resol}$ and $M_\mathrm{Resol}$ columns).  According to our results, CHF outperforms TopoCluster on all three datasets. Moreover, \citeauthor{DeLozier:2015:GTR:2886521.2886652} \cite{DeLozier:2015:GTR:2886521.2886652} set the error distance threshold for TopoCluster to 161 kilometers and achieved an accuracy of $71.4\%$ on $LGL$\footnote{Among the datasets used in TopoCluster paper \cite{DeLozier:2015:GTR:2886521.2886652}, $LGL$ is the only dataset to which we have access}, whereas under the same setting, CHF reaches $71.2\%$ on $LGL$, which is marginally lower than TopoCluster.

Besides accuracy, the mean error distance is also measured in our {\em Resol} experiments\footnote{Mean error distance for TopoCluster in $LGL$ is derived from the original paper \cite{DeLozier:2015:GTR:2886521.2886652}.}. Among the unsupervised methods, CBH stands out with the lowest error. CHF is close to CBH with its error not exceeding 40km.
This difference stems from SHS impacting CHF because when a toponym is projected to an incorrect location by SHS, the mapped location is more likely located in a country different than the ground truth.

\subsection{Unseen Data Analysis}
\begin{figure}[t!]
\includegraphics[width=2.82in]{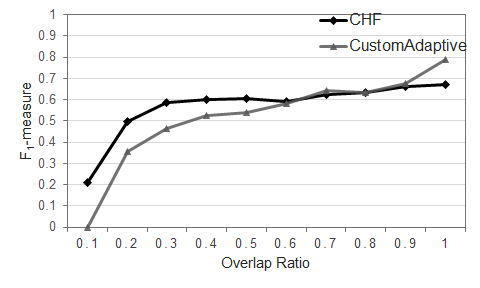}
\caption{$F_1$-$measure$ of CustomAdaptive trained on CLUST and CHF when overlap ratio varies. CHF yields a better performance than CustomAdaptive when overlap between training data and test data is lower than 60\%.}
\label{fig:OverlapStudy}
\end{figure}

Supervised techniques benefit from the knowledge gained in the training phase and if there is an overlap between the training data and the test data, then the prediction can be counted as overly optimistic. \citeauthor{Domingos:2012:FUT:2347736.2347755} emphasizes that generalization is achieved by a separation of the training data and the test data \cite{Domingos:2012:FUT:2347736.2347755}. This is why, we study the effect of the overlap between training and test datasets on $F_1$-$measure$. For this purpose, CustomAdaptive classifier was trained on $CLUST$ dataset (the trend does not vary significantly if the classifier trained on $LGL$) and tested against $TR$-$News$. We define the overlap ratio measure as the number of toponyms per article in test data, which has also been appeared in the training data. We can channel overlap ratio through trimming off articles from test data and measure performance on the trimmed test data. Figure \ref{fig:OverlapStudy} plots $F_1$-$measure$ against the overlap ratio. The unsupervised method surpasses the supervised method when the overlap ratio is less than $60\%$ (when the overlap ratio is at 0.6, CHF still outperforms CustomAdaptive with a 1\% margin). This observation confirms that the unsupervised technique, namely CHF, can handle unknown data better than the supervised method, namely Adaptive (CustomAdaptive 
implementation).

\section{Related Works}
\label{sec:RelatedWorks}
Numerous studies have been conducted and much progress has been made on the task of disambiguating location mentions. The existing approaches in the literature may be grouped into (1) unsupervised and rule-based, (2) supervised, and (3) those based on some knowledge bases.   
However, a plethora of methods leverage a mixture of techniques. For example, \citeauthor{DeLozier:2015:GTR:2886521.2886652} \cite{DeLozier:2015:GTR:2886521.2886652} proposed an unsupervised toponym resolution method that leverages geographical kernels and spatially annotated Wikipedia articles. Also, \citeauthor{Lieberman:2012:ACF:2348283.2348381} \cite{Lieberman:2012:ACF:2348283.2348381} presented a supervised technique that uses both geographical distance and additional knowledge like gazetteers and document source to disambiguate toponyms.


\begin{table*}[t]
\begin{center}
 \begin{tabular}{l|c c c c c|c c c c c|c c c c c} 
 \hline
  & \multicolumn{5}{c|}{\em LGL} & \multicolumn{5}{c|}{\em CLUST} & \multicolumn{5}{c}{\em TR-News} \\
  \textbf{Method} & $P$ & $R$ & $F_1$ & $P_\mathrm{Resol}$ & $M_\mathrm{Resol}$ & $P$ & $R$ & $F_1$ & $P_\mathrm{Resol}$ & $M_\mathrm{Resol}$ & $P$ & $R$ & $F_1$ & $P_\mathrm{Resol}$ & $M_\mathrm{Resol}$ \\
 \hline
 \textbf{Unsupervised} & & & & & & & & & & & & & & & \\
 CBH & 66.8 & 40.6 & 50.5 & 68.6 & \textbf{760}    & \textbf{80.6} & 55.8 & 66.0 & \textbf{81.5} & \textbf{709}   & 74.9 & 53.0 & 62.1 & 79.2 & \textbf{869} \\
 SHS & \textbf{69.7} &	\textbf{43.3} & \textbf{53.4} & 68.3 & 1372    & 72.8 &	51.6 &	60.4 & 71.1 & 1521   & 73.8 & 53.6 & 62.1 & 69.9 &  2305 \\
 CHF & 68.5 & 43.1 & 52.9 & \textbf{68.9} & 818    & \textbf{80.6} & \textbf{58.4} & \textbf{67.7} & 81.0 & 788   & \textbf{79.3} & \textbf{58.2} &	\textbf{67.1} & \textbf{80.5} & 942 \\
 TopoCluster \cite{DeLozier:2015:GTR:2886521.2886652}   & - & - & - & 59.7 & 1228    & - & - & - & 77.1 & 769     & - & - & - & 68.8 & 1422 \\
 \hline
 \textbf{Supervised} & & & & & & & & & & & & & & &\\
 Adaptive \cite{Lieberman:2012:ACF:2348283.2348381} & - & \textbf{58.7} & - & \textbf{94.2} & -    & - & \textbf{61.8} & - & \textbf{96.0} & -   & - & - & - & - & - \\ 
 CustomAdaptive & \textbf{79.2} & 48.5 & \textbf{60.2} & 88.3 & \textbf{679}   & \textbf{89.8} & 57.9 & \textbf{70.4} & 93.4 & \textbf{504}   & \textbf{83.8} &	\textbf{74.9} & \textbf{79.1} & \textbf{90.5} & \textbf{573} \\
 \hline
 \textbf{Commercial} & & & & & & & & & & & & & & & \\
 Placemaker & 73.5 & \textbf{48.6} & \textbf{58.5} & - & -   & 87.4 & \textbf{61.1} & \textbf{71.9} & - & -   & 80.8 & \textbf{63.0} & \textbf{70.8}  & - & - \\
 OpenCalais & 77.1 & 28.9 & 42.1 & - & -    & \textbf{87.5} & 48.5 & 62.4 & - & -   & \textbf{81.3} & 48.5 & 61.2 & - & - \\
 GoogleNL+Wiki & \textbf{80.5} & 34.0 & 47.8 & - & -   & 82.8 & 39.2 & 53.2 & - & -   & 80.2 & 38.4 & 51.9 & - & -   \\
 \hline
\end{tabular}
\caption{Performance results of various methods in {\em GeoTag} and {\em Resol} experiments. The best results in each category are bolded.}
\label{tab:resolution}
\end{center}
\end{table*}

\noindent
{\bf Unsupervised and rule-based methods}
In unsupervised 
resolution, various techniques have been studied. Map-based methods create a representation of all referents on a world map and apply techniques such as 
geographical centroid detection and outlier elimination to estimate the target of a toponym \cite{Leidner:2007:TRT:1328964.1328989}. \citeauthor{Moncla:2014:GTF:2666310.2666386} \cite{Moncla:2014:GTF:2666310.2666386} introduce a map-based technique where density-based clustering was carried out to detect outliers. \citeauthor{Buscaldi:2011:ADT:2047296.2047300} \cite{Buscaldi:2011:ADT:2047296.2047300} argues that map-based techniques face difficulties in grounding toponyms in a document when they are spatially far from each other.
Rule-based and heuristic-based methods also have been adopted in the literature \cite{Leidner:2007:TRT:1328964.1328989, Amitay:2004:WGW:1008992.1009040}. For instance, the presence of ``Canada'' in text \textit{London, Canada} may help disambiguate \textit{London}. However, finding a set of rules to cover all cases in natural language text seems to be arduous. 

\noindent
{\bf Approaches using knowledge bases}
Wikipedia has been integrated as a knowledge base into more recent toponym disambiguation techniques
 \cite{Ardanuy:2017:TDH:3078081.3078099,DeLozier:2015:GTR:2886521.2886652,Spitz:2016:SFA:2948649.2948651, speriosu-baldridge:2013:ACL2013, Santos2015}.
\citeauthor{Ardanuy:2017:TDH:3078081.3078099} \cite{Ardanuy:2017:TDH:3078081.3078099} address toponym disambiguation in multilingual retrospective articles. They build a model to distill semantic features from Wikipedia information such as page title and article body. \citeauthor{speriosu-baldridge:2013:ACL2013} \cite{speriosu-baldridge:2013:ACL2013} argue that non-spatial words impart useful information to disambiguate toponyms and they propose likelihood models that are obtained from Wikipedia. \citeauthor{DeLozier:2015:GTR:2886521.2886652} \cite{DeLozier:2015:GTR:2886521.2886652} propose TopoCluster, which does not rely on gazetteers to resolve toponyms, to address cases where location mentions are not found in gazetteers. They construct a geographical language model to capture geographical senses of words using Wikipedia pages of locations. However, they note that adding gazetteer information to TopoCluster, namely \textit{TopoClusterGaz}, yields a better performance. Less known toponyms are not expected to be found in Wikipedia; they can introduce challenges and hinder the performance of this method. 

\noindent
{\bf Supervised methods}
Many classification techniques have been proposed for geotagging purposes including Bayesian \cite{Adelfio:2013:GTR:2525314.2525321}, random forests \cite{Lieberman:2012:ACF:2348283.2348381}, RIPPER rule learner \cite{garbin-mani:2005:HLTEMNLP} and SVM \cite{garbin-mani:2005:HLTEMNLP, Melo:2015:GTD:2837689.2837690}. The features extracted for these classifiers can be grouped into context-free and context-sensitive features \cite{Lieberman:2012:ACF:2348283.2348381}. 
Context-free features typically include heuristics and information from external sources such as knowledge bases and gazetteers and may include, for example, population \cite{Lieberman:2012:ACF:2348283.2348381} and location type \cite{garbin-mani:2005:HLTEMNLP}. Context-sensitive features are obtained from documents where toponyms are mentioned. \citeauthor{Melo:2015:GTD:2837689.2837690} \cite{Melo:2015:GTD:2837689.2837690} use normalized TF-IDF document vectors over curvilinear and quadrilateral regions on Earth's surface. The adaptive method, proposed by
\citeauthor{Lieberman:2012:ACF:2348283.2348381}
\cite{Lieberman:2012:ACF:2348283.2348381}, casts geographical proximity and sibling relationship among interpretations in a context window as features. GeoWhiz \cite{Adelfio:2013:GTR:2525314.2525321} aggregate several likelihoods based on observations in training data. For instance, largely populated places are more likely estimated as their prominent interpretation. The suggested method by \citeauthor{Santos2015} \cite{Santos2015} consolidates information from Wikipage pages of locations to compute several similarity and geographical features (context-free features) and performs a nearest neighbor search using locality-sensitive hashing (LSH) to resolve locations.

\noindent
{\bf Other more general related work}
Entity disambiguation (also known as entity linking) \cite{Li:2013:MEN:2487575.2487681, TACL528, 6823700, Hoffart:2014:DEE:2566486.2568003, Ganea:2016:PBM:2872427.2882988} is related to toponym resolution. Linking named entities (i.e., people, organizations, and locations) to their corresponding real world entities in a knowledge base subsumes toponym disambiguation. Nonetheless, geographical features of location entities are neglected by these systems \cite{speriosu-baldridge:2013:ACL2013} and thus, geographically specialized methods for resolving toponyms are still needed to map locations to their corresponding geographic footprint.

Another line of research pertinent to this work is location disambiguation in social media. The related work in this area may incorporate user profile data and social network information as well as natural language processing tools and gazetteers to tackle this task~\cite{Ikawa:2013:LIS:2487788.2488107,Muthiah:2015:PPM:2888116.2888259}.
\citeauthor{Flatow:2015:AHG:2684822.2685296} \cite{Flatow:2015:AHG:2684822.2685296} propose a method that learns geo-referenced $n$-grams from training data to perform geotagging on social messages.
Use of words that are endogenous to social media are considered as an inherent hurdle here. Moreover, social media content have deficient orthographic structure and lack context, which bring even more complexities to toponym resolution in social media \cite{eisenstein:2013:LASM, Ritter:2012:ODE:2339530.2339704, matsuda-EtAl:2015:LAW}.

\section{Conclusions}
\label{sec:Conclusion}
In this paper, we study toponym resolution and propose two novel unsupervised models and a mixture model, namely CHF, to address the problem. We investigate the effectiveness of the proposed methods with other techniques. Our evaluations show that the Context-Hierarchy Fusion method outperforms TopoCluster, the state-of-the-art unsupervised method, in terms of precision.
The performance of supervised techniques exceeds that of our proposed methods (as expected), nonetheless, we have shown that the state-of-the-art supervised classifier, called Adaptive, highly relies on the training data and Context-Hierarchy Fusion can handle unseen toponyms better.

The future work may investigate other mixture models and a better understanding of when one or both of supervised and unsupervised methods are expected to perform not so well. 
In addition, the correlations among the bounding-boxes of toponyms in an article can be studied to augment the resolution, considering the gazetteer are endowed with bounding-box of locations for this purpose \cite{KumarSingh2018}. Another direction is understanding the differences between short and long text as far as toponym resolution is concerned and the challenges each pose.

\begin{acks}
This research is supported by the Natural Sciences and Engineering Research Council of Canada.
The authors are grateful to Michael Lieberman for sharing the $LGL$ and $CLUST$ datasets.
The authors would also like to thank the anonymous referees for their valuable comments and helpful suggestions.
\end{acks}

\bibliographystyle{ACM-Reference-Format}
\bibliography{paper-bibliography} 

\end{document}